\documentstyle[prl,aps,psfig,preprint]{revtex}
\input{epsf}                                                               
\tightenlines
\begin{document}                                                           
                                                                           
                                                                           
\draft
                                                                           
\title{\bf {Search for the pentaquark via the                              
$P^0_{{\bar c}s} \rightarrow \phi\pi p$ decay}}

                                                                           
\author{                                                                   
    E.~M.~Aitala,$^9$                                                      
       S.~Amato,$^1$                                                       
    J.~C.~Anjos,$^1$                                                       
    J.~A.~Appel,$^5$                                                       
       D.~Ashery,$^{15}$                                                   
       S.~Banerjee,$^5$                                                    
       I.~Bediaga,$^1$                                                     
       G.~Blaylock,$^8$                                                    
    S.~B.~Bracker,$^{16}$                                                  
    P.~R.~Burchat,$^{14}$                                                  
    R.~A.~Burnstein,$^6$                                                   
       T.~Carter,$^5$                                                      
 H.~S.~Carvalho,$^{1}$                                                     
    N.~K.~Copty,$^{13}$                                                    
    L.~M.~Cremaldi,$^9$                                                    
 C.~Darling,$^{19}$                                                        
       K.~Denisenko,$^5$                                                   
       A.~Fernandez,$^{12}$                                                
       P.~Gagnon,$^2$                                                      
       S.~Gerzon,$^{15}$                                                   
       C.~Gobel,$^1$                                                       
       K.~Gounder,$^9$                                                     
     A.~M.~Halling,$^5$                                                    
       G.~Herrera,$^4$                                                     
 G.~Hurvits,$^{15}$                                                        
       C.~James,$^5$                                                       
    P.~A.~Kasper,$^6$                                                      
       S.~Kwan,$^5$                                                        
    D.~C.~Langs,$^{11}$                                                    
       J.~Leslie,$^2$                                                      
       J.~Lichtenstadt,$^{15}$                                             
       B.~Lundberg,$^5$                                                    
       S.~MayTal-Beck,$^{15}$                                              
       B.~Meadows,$^3$                                                     
 J.~R.~T.~de~Mello~Neto,$^1$                                               
    R.~H.~Milburn,$^{17}$                                                  
 J.~M.~de~Miranda,$^1$                                                     
       A.~Napier,$^{17}$                                                   
       A.~Nguyen,$^7$                                                      
  A.~B.~d'Oliveira,$^{3,12}$                                               
       K.~O'Shaughnessy,$^2$                                               
    K.~C.~Peng,$^6$                                                        
    L.~P.~Perera,$^3$                                                      
    M.~V.~Purohit,$^{13}$                                                  
       B.~Quinn,$^9$                                                       
       S.~Radeztsky,$^{18}$                                                
       A.~Rafatian,$^9$                                                    
    N.~W.~Reay,$^7$                                                        
    J.~J.~Reidy,$^9$                                                       
    A.~C.~dos Reis,$^1$                                                    
    H.~A.~Rubin,$^6$                                                       
 A.~K.~S.~Santha,$^3$                                                      
 A.~F.~S.~Santoro,$^1$                                                     
       A.~J.~Schwartz,$^{11}$                                              
       M.~Sheaff,$^{4,18}$                                                   
    R.~A.~Sidwell,$^7$                                                     
    A.~J.~Slaughter,$^{19}$                                                
    M.~D.~Sokoloff,$^3$                                                    
       N.~R.~Stanton,$^7$                                                  
       K.~Stenson$^{18}$                                                   
    D.~J.~Summers,$^9$                                                     
 S.~Takach,$^{19}$                                                         
       K.~Thorne,$^5$                                                      
    A.~K.~Tripathi,$^{10}$                                                 
       S.~Watanabe,$^{18}$                                                 
 R.~Weiss-Babai,$^{15}$                                                    
       J.~Wiener,$^{11}$                                                   
       N.~Witchey,$^7$                                                     
       E.~Wolin,$^{19}$                                                    
       D.~Yi,$^9$                                                          
       S.~Yoshida,$^7$                                                     
       R.~Zaliznyak,$^{14}$                                                
       and                                                                 
       C.~Zhang$^7$ \\                                                     
\begin{center} (Fermilab E791 Collaboration) \end{center}                  
}

\address{                                                                  
$^1$ Centro Brasileiro de Pesquisas F\'isicas, Rio de Janeiro, Brazil\\        
$^2$ University of California, Santa Cruz, California 95064\\              
$^3$ University of Cincinnati, Cincinnati, Ohio 45221\\                    
$^4$ CINVESTAV, Mexico\\                                                   
$^5$ Fermilab, Batavia, Illinois 60510\\                                   
$^6$ Illinois Institute of Technology, Chicago, Illinois 60616\\           
$^7$ Kansas State University, Manhattan, Kansas 66506\\                    
$^8$ University of Massachusetts, Amherst, Massachusetts 01003\\           
$^9$ University of Mississippi, University, Mississippi 38677\\            
$^{10}$ The Ohio State University, Columbus, Ohio 43210\\                  
$^{11}$ Princeton University, Princeton, New Jersey 08544\\                
$^{12}$ Universidad Autonoma de Puebla, Mexico\\                           
$^{13}$ University of South Carolina, Columbia, South Carolina 29208\\     
$^{14}$ Stanford University, Stanford, California 94305\\                  
$^{15}$ Tel Aviv University, Tel Aviv, Israel\\                            
$^{16}$ 317 Belsize Drive, Toronto, Canada\\                               
$^{17}$ Tufts University, Medford, Massachusetts 02155\\                   
$^{18}$ University of Wisconsin, Madison, Wisconsin 53706\\                
$^{19}$ Yale University, New Haven, Connecticut 06511\\                    
}                                                                          
                                                                           
\maketitle

\setlength{\baselineskip}{2.5ex}                                            
                                                                           
\begin{abstract}                                                           
We report results of the first search for the pentaquark                   
$P_{{\bar c}s}$ which is predicted to be a doublet of states:              
$P^0_{{\bar c}s}=|\bar{c} s u u d\rangle$ and                              
$P^-_{{\bar c}s}=|\bar{c} s d d u\rangle$. A search was made for the decay 
$P^0_{{\bar c}s} \rightarrow \phi \pi p$ in data from                      
Fermilab experiment E791, in which                                         
500~GeV/$c$ $\pi^-$ beam interacted with nuclear targets.                  
We present upper limits at 90$\%$ confidence level for the ratio of        
cross section times branching fraction of this decay                       
to that for the decay $D_s^{\pm} \rightarrow \phi\pi^{\pm}$.               
The upper limits are 0.031 and 0.063 for $M(P^0_{{\bar c}s})$ = 2.75 and   
2.86 GeV/$c^2$,                                                            
respectively, assuming a $P^0_{{\bar c}s}$ lifetime of 0.4 ps.             
\end{abstract}                                                             
                                                                           
                                                                           
\narrowtext                                                                
                                                                           
\noindent                                                                  
The spectrum of observed hadrons fits into multiplets of two- and 
three-quark states. The mass differences within these multiplets can        
be explained by effective quark masses and the color-hyperfine (CH)         
interaction in the QCD Hamiltonian.                                         
Calculations done using the CH interaction predict                          
the existence of particles made of more than three quarks.                  
Jaffe \cite{jaf} predicted the existence of the $H$ dibaryon,               
$H = |u u d d s s\rangle$,                                                  
and extensive efforts have been made to find it experimentally \cite{alan}. 
Lipkin \cite{lip} and  Gignoux {\em et al.} \cite{rich}                     
have proposed that a doublet of states,                                     
the $P^0_{{\bar c}s}=|\bar{c} s u u d\rangle$ and the                       
$P^-_{{\bar c}s}=|\bar{c} s d d u\rangle$, and                              
their charge conjugate states, may exist and be stable against              
strong decays. These                                                        
were named pentaquarks.\\                                                   
                                                                            
\noindent                                                                   
The threshold for strong decay of the pentaquark is 2907 MeV/$c^2$, above   
which it can decay to $D_s^{\pm}$ and a nucleon.                            
Calculations done using only the CH interaction predict pentaquark masses   
which vary from 150 MeV/$c^2$ below the                                     
$D^{\pm}_s$--nucleon threshold to a few tens of MeV/$c^2$ below threshold,  
depending on how SU(3)$_{\it flavor}$ symmetry breaking and the mass of the 
charm antiquark are taken into account \cite{rich}. Contributions to the    
binding energy from other components of the Hamiltonian are even more model 
dependent.                                                                 
Calculations done using an Instanton model \cite{tnk},                     
bag models \cite{zr,fle} and a Skyrme model \cite{rs}                      
conclude that, depending upon the choice of parameters, the                
pentaquark is bound or is a near-threshold resonance.                      
If bound, the lifetime is expected to be similar to that of charm particles,
with the exact value depending upon unknown internal structure.             
In a description of the pentaquark as                                       
an off-shell charm meson and a spectator baryon,                            
it is assumed that the off-shell meson                                      
decays to the same decay products as the free meson.                        
The pentaquark lifetime should then be similar to that of the $D_s^{\pm}$   
charm meson, 0.47 ps.                                                       
A description of the pentaquark as a five-quark state                       
allows more interactions among the quarks and                               
consequently may lead to a shorter lifetime.                                
In the work described here, we have considered lifetimes ranging from 0.1 to
1.0 ps, and pentaquark masses between 2.75 and 2.91 GeV/$c^2$.\\

\noindent                                                                   
Various mechanisms for pentaquark production have been discussed by         
H. Lipkin~\cite{lipkin2}.                                                  
However, only crude estimates of the pentaquark production cross-section   
exist in the literature.                                                   
One mechanism considers a                                                  
production of all five quarks in the interaction \cite{dpf94} and          
is based on an empirically motivated equation which predicts reasonably    
well the production cross-section of other charm particles.                
Another mechanism is the coalescence                                       
model, where pentaquark components such as the $D_s^{\pm}$ and a nucleon are 
produced in the reaction and                                                
fuse into one particle while in overlapping regions of                      
phase-space \cite{danmur}.                                                  
Typically, the estimated pentaquark production cross-section is             
of the order of 1$\%$                                                       
that of the $D_s^{\pm}$.\\

\noindent                                                                   
In this letter we report results from the first search for                  
$P^0_{{\bar c}s}$ production, which was carried out in experiment           
E791 at Fermilab.                                                           
We measure the product of cross section and branching fraction 
($\sigma\cdot$B)
for $P^0_{{\bar c} s}\rightarrow \phi\pi p \rightarrow K^+K^-\pi p$         
relative to that for                                                        
$D^{\pm}_s\rightarrow \phi\pi^{\pm} \rightarrow K^+K^-\pi^{\pm}$.\\

\noindent                                                                   
The E791 experiment \cite{e791} recorded $2\times 10^{10}$ events from      
interactions of a 500 GeV/c $\pi^-$ beam in five thin targets               
(one platinum, four diamond) separated by gaps of 1.34 to 1.39 cm.          
The trigger included a loose requirement on transverse energy deposited     
in the calorimeters by particles coming from the interaction.               
Precision vertex and tracking information was provided by 23 silicon        
microstrip detectors (6 upstream of the targets and 17 downstream),         
ten proportional wire-chamber planes (8 upstream, 2 downstream), and        
35 drift-chamber planes. Momentum was measured                              
using two dipole magnets.                                                   
Two multicell, threshold \v{C}erenkov counters were used for $\pi$, $K$ and 
$p$ identification.                                                         
The velocity threshold for light emission was different in the two          
\v{C}erenkov counters, allowing for                                         
discrimination of charged hadrons in                                        
certain momentum ranges.                                                    
The amount of light collected for each reconstructed particle trajectory    
was compared to that expected for each                                      
mass hypothesis for the measured momentum. This comparison resulted         
in a probability estimate for each type of charged hadron~\cite{cer}.       
Selection criteria                                                          
based on the probability values were used to identify the particles.        
\\

\noindent                                                                   
We have searched for the pentaquark in its expected decay mode              
$P^0_{{\bar c}s}\rightarrow\phi\pi p$, where the $\phi$ subsequently        
decays to $K^+K^-$. We normalize the sensitivity of our search to           
$D_s^{\pm}\rightarrow \phi\pi^{\pm}$ decays which are similar enough that   
several systematic errors are common to both decay modes and cancel         
in the measured ratio of cross sections and branching fractions.            
These are convenient decay                                                  
modes to detect because all decay products are charged,                     
and because the narrow $\phi$                                               
signal allows a rejection of $K^+K^-$ background.                           
We calculate the acceptance of our detector via Monte Carlo simulation.     
The production of the pentaquark was simulated using the PYTHIA             
particle generator \cite{lund}. The pentaquark was                          
introduced into the LUND list                                               
of particles (replacing the $\Xi_c^0$) and was forced to decay to           
$\phi\pi p\rightarrow K^+ K^- \pi p$. The pentaquark was simulated          
with a lifetime of 0.4 ps and with                                          
two different masses, 2.75 and 2.86 GeV/$c^2$.\\                            
                                                                            
\noindent                                                                   
The first stage of data analysis included reconstruction of tracks and      
vertices. Events with evidence of multiple vertices were kept for further   
analysis.                                                                   
In the pentaquark analysis, we searched for events with four tracks         
emerging from a decay vertex,                                               
consistent with being $K\; K\; \pi\; p$ according to information            
from the \v{C}erenkov counters.                                             
It was required that the two kaons have opposite                            
charge and that the total charge of the four tracks be zero. The invariant  
mass of the two kaons                                                       
was required to be within $\pm 5$ MeV/$c^2$ of the $\phi$ mass.             
This process selects $P^0_{{\bar c}s}$ and ${\bar P^0_{c{\bar s}}}$         
candidates with equal sensitivity.                                          
Topological, kinematic, and other criteria were imposed to reject background
and improve the statistical significance of a pentaquark                    
signal                                                                      
in the $\phi\pi p$ mass spectrum.                                           
To optimize the selection criteria with minimum bias, a sensitivity         
function $S/\sqrt{BG}$ was defined where $S$ is the number of signal        
events and $BG$ is the number of background events that would pass          
the selection criterion. This function was plotted versus the               
value of selection criteria applied to the discrimination variable          
of interest (track and vertex goodness of fit, {\v C}erenkov                
identification probabilities, kinematical data, {\it etc.}).                
Two models of signal were used to determine the yield $S$.                  
One was $P^0_{{\bar c}s}$ generated with the Monte Carlo                    
simulation. The other was the signal from the decay of a known particle,    
present in the data and having certain properties similar                   
to those expected for the pentaquark decay. We used the                     
$D^0\rightarrow K\pi\pi\pi$ decay to optimize four-prong topological 
criteria. A sample of $\phi\rightarrow K^+K^-$ decays,                      
selected independently from the                                             
$\phi\pi p$ sample, was used to optimize \v{C}erenkov                       
kaon identification criteria.                                               
The number of background events, $BG$, was determined from the $\phi\pi p$  
invariant mass spectrum,                                                    
in a mass region outside the 2.75 to 2.91 GeV/$c^2$ range.                  
The maximum in the sensitivity function determined                          
the selection criterion for each discrimination variable.                   
By comparing the results using the two sources                              
of signal, we                                                               
selected the criteria such that the sensitivity function was                
stable against small changes in criterion values. When this procedure       
resulted in a range of acceptable criteria, we selected the criterion that  
yielded the highest efficiency.\\                                           
                                                                            
\noindent                                                                   
The resulting optimal kaon identification criteria were that each of        
the two tracks have                                                         
momentum between 6 and 40 GeV/$c$ and a certain minimum \v{C}erenkov        
identification probability. These conditions excluded more than 85$\%$ of   
pions                                                                       
and 60$\%$ of protons, while accepting about 70$\%$ of kaons.               
The pion {\v C}erenkov identification                                       
requirement excluded about 75$\%$ of protons and kaons, while accepting     
about 70$\%$ of pions.                                                      
The proton identification requirement excluded more than 35$\%$ of pions,   
while accepting more than 90$\%$ of protons and kaons.\\                    
                                                                            
\noindent                                                                   
The topological criteria included requirements on the quality               
of the reconstructed production and decay vertices ($\chi^2$ of the fit),   
a separation between the vertices of                                        
more than 10$\sigma_\ell$ where $\sigma_\ell$ is the error in the measured  
longitudinal separation, and a demand that                                  
the momentum vector of the candidate pentaquark                             
points back to the production vertex with an impact parameter of less than  
25 $\mu$m. Each of the four decay tracks was required to pass at least twice 
as close                                                                    
to the decay vertex as to the production vertex. In addition, the product of
the ratios                                                                  
of the distance of closest approach of these tracks to the decay vertex and 
to the production vertex was required to be less than 0.001. The decay 
vertex had to be         
isolated from its neighboring tracks by at least 10 $\mu$m and to be        
located significantly outside any of the target foils. If one of the        
tracks emerging                                                             
from the candidate decay vertex pointed to another vertex located           
within one of the downstream targets, it was rejected as possibly arising   
from a secondary interaction.                                               
A selection criterion was applied on the sum of the squared transverse 
momenta 
(p$_t^2$) of the four tracks, relative to the direction of their            
summed momentum, $\sum$p$_t^2 > $0.5 (GeV/$c$)$^2$.                        
Since this quantity                                                        
is determined by the Q value of the decay (about 800 MeV/$c$               
for the pentaquark) the optimization procedure for $\sum$p$_t^2$ was       
done using only the signal from the Monte Carlo simulation, which was 
found to
reproduce well the distribution of this parameter for known decays.\\      
                                                                           
\noindent                                                                  
Other background reduction criteria included elimination of                
$\phi\pi p$ vertices that contain either $\Lambda\rightarrow\pi p$         
candidates or a $\phi$ that points back within 45 $\mu$m of the production 
vertex.                                                                    
Known particles decaying to four-prongs with the decay products            
(mis)identified as two kaons, a pion and a proton could appear in the      
$\phi\pi p$ invariant mass spectrum as a flat background or as a broad peak.
We identified no known particles that form a peak in the $\phi\pi p$        
mass window and the only source for                                         
flat background due to misidentification of the tracks was from             
the decay $D^0\rightarrow K\pi\pi\pi$.                                      
Candidate $K K \pi p$ events with a $K\pi\pi\pi$ invariant mass consistent  
with the $D^0$ mass were removed.                                           
Above the                                                                   
appropriate thresholds, candidate $\phi\pi p$ events could be due to the    
combinations $(\Lambda_c\rightarrow\phi p)\; +\; \pi$ or                    
$(D^{\pm}_s,D^{\pm}\rightarrow\phi\pi^{\pm})\; +\; p$.                      
No $\Lambda_c\rightarrow\phi p$ candidates were found within the            
$\phi\pi p$ sample, but three events for which the $\phi\pi$ invariant      
mass is consistent with the $D^{\pm}_s$ or D$^{\pm}$ masses passed all the 
analysis cuts.\\                                                            
                                                                            
\begin{figure}[htbp]
\centerline{\epsfxsize=8cm \epsfbox{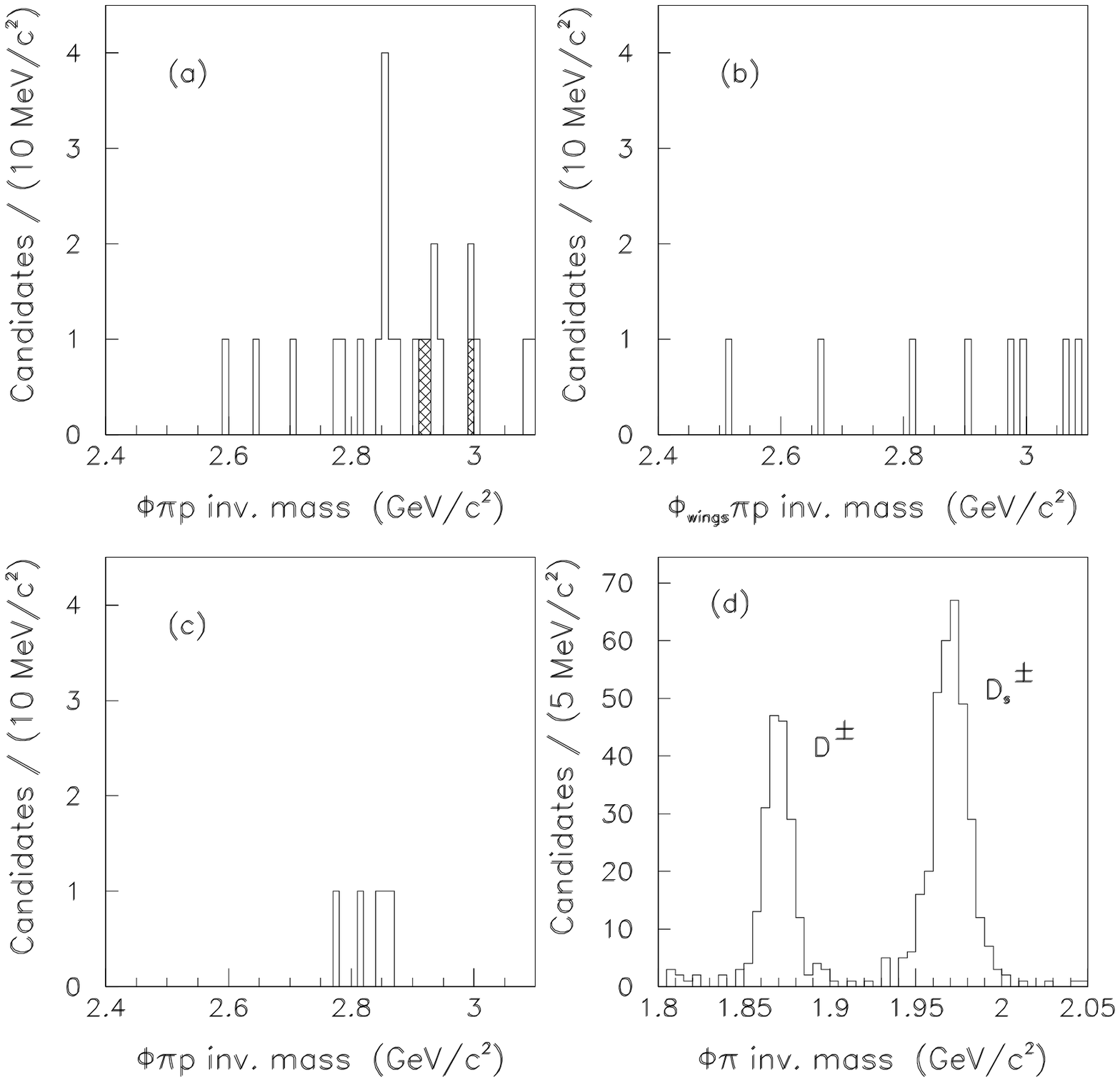} }
\caption{  
(a) The $\phi\pi p$ invariant mass spectrum from the 
E791 data for the optimized selection criteria.
Events in which the $\phi\pi$ invariant mass is consistent with the 
D$^{\pm}$ or $D^{\pm}_s$ masses are shaded.
(b) Spectrum of $\phi_{ wings}\pi p$ for the optimized selection 
criteria; see text for a full description.   
(c) The same spectrum as in (a), with a tighter proton identification 
criterion. 
(d) $\phi\pi$ invariant mass spectrum showing the $D^{\pm}_s$ 
normalization sample. The left-most peak arises from Cabibbo-suppressed
$D^{\pm}$ decays.}
\label{fig:spectra}
\end{figure}

\noindent                                                                   
In Fig.~\ref{fig:spectra}(a) we show the final $\phi\pi p$                  
invariant mass spectrum for the optimized analysis cuts. The three events   
which could be described as 
$(D^{\pm}_s,D^{\pm}\rightarrow\phi\pi^{\pm})\; +\; p$ are 
shaded. In Fig.~\ref{fig:spectra}(b) we show the                            
$\phi_{wings}\pi p$ invariant mass spectrum, where $\phi_{wings}$ refers    
to $K^+K^-$ candidates with invariant mass in a range {\em outside} the 
required $\phi$ mass window (between 5 and 10 MeV/$c^2$ below and above 
the $\phi$ mass).
This spectrum contains essentially only background events. 
In Fig.~\ref{fig:spectra}(c) we show the $\phi\pi p$ invariant mass        
spectrum with a                                                            
tighter cut on the proton \v{C}erenkov probability, together with          
a requirement that the track momentum be between                           
22 and 75 GeV/$c$. This selection criterion excludes                        
95$\%$ of pions and more than 80$\%$ of kaons.                              
It gives essentially the same                                               
sensitivity for a pentaquark signal but                                     
with half the efficiency.                                                   
In Fig.~\ref{fig:spectra}(d) we show the $\phi\pi$ invariant mass           
spectrum for the $D^{\pm}_s\rightarrow\phi\pi^{\pm}$ normalization sample.  
This sample was selected using the same selection criteria (where relevant) 
as were used to select pentaquark candidates. In this manner the systematic 
error on the ratio of efficiencies for the two decay modes was minimized.\\

\noindent                                                                   
The $\phi\pi p$ invariant mass spectrum (Fig.~\ref{fig:spectra}(a))         
shows a concentration of seven events near 2.86 GeV/$c^2$ which is          
absent in the background spectrum of Fig.~\ref{fig:spectra}(b).             
Three of these seven events                                                 
survive the tighter proton \v{C}erenkov selection criterion described above 
(Fig.~\ref{fig:spectra}(c)), consistent with the expected efficiency of     
this criterion.                                                             
Only two events outside the concentration                                   
survive this requirement.                                                   
On the other hand, the proton candidate tracks project back to the 
production
vertex with an impact parameter distribution different from that predicted  
by the Monte Carlo simulation. For that reason (given the number of         
events in our final sample),                                               
we conclude that there is no                                               
evidence for $P^0_{{\bar c}s}\rightarrow\phi \pi p$ decays in our data.\\

\noindent                                                                  
We use the spectrum of Fig.~\ref{fig:spectra}(a) to obtain 90\% C.L.\      
upper limits on                                                            
the product of the pentaquark production cross section and the pentaquark  
branching fraction to $\phi\pi p$, relative to that for                    
$D_s^{\pm}\rightarrow\phi\pi^{\pm}$.                                       
For a particular $\phi\pi p$ invariant mass, our limit is:                 
\begin{equation}                                                           
\label{eqn:ratio}                                                          
UL\left(\frac{\sigma^{}_P\cdot B^{}_{P\rightarrow\phi\pi p}}               
{\sigma^{}_{D^{}_s}\cdot B^{}_{D^{}_s\rightarrow\phi\pi}}\right) =         
\frac{UL(N^{}_{P\rightarrow\phi\pi p})/
      \varepsilon^{}_{P\rightarrow\phi\pi p}}
     {N^{}_{D^{}_s\rightarrow\phi\pi}/
     \varepsilon^{}_{D^{}_s\rightarrow\phi \pi}},
\end{equation}                       
where $UL(N^{}_{P\rightarrow\phi\pi p})$ is the 90\% C.L.\ upper limit     
on the number of signal events in a mass window centered on                
the invariant mass of interest, given the number of events                 
observed in the window and the expected number of background               
events\cite{pdg}. The quantity $N^{}_{D^{}_s\rightarrow\phi\pi}$ is the    
number of $D^{\pm}_s\rightarrow\phi\pi^{\pm}$ decays obtained              
from the normalization sample (Fig.~\ref{fig:spectra}d), and the           
quantities $\varepsilon^{}_{P\rightarrow\phi \pi p}$                       
and $\varepsilon^{}_{D^{}_s\rightarrow\phi \pi}$ are                       
the detection efficiencies for                                             
$P^0_{{\bar c}s}\rightarrow\phi \pi p$ and 
$D^{\pm}_s\rightarrow\phi\pi^{\pm}$, respectively.
These efficiencies were calculated from Monte Carlo simulation.            
The $P\rightarrow\phi\pi p$ efficiency depends on the pentaquark mass      
and for $M(P^0_{{\bar c}s}) \simeq 2.86$ GeV/$c^2$ is approximately 
0.13$\%$. The shape of the background spectrum ---\,for lack of more 
information\,--- 
is assumed to be flat as indicated by Fig.~\ref{fig:spectra}(b).
The level of background expected is obtained from Fig.~\ref{fig:spectra}(a) 
by matching the flat spectrum to the number of events                       
between 2.4 and 3.1~GeV/$c^2$, excluding                                    
the shaded events and                                                       
the mass region between 2.73 to 2.91 GeV/$c^2$.\\                           
                                                                            
\noindent                                                                   
In this paper we present limits for two different pentaquark                
masses: 2.75~GeV/$c^2$ and 2.86~GeV/$c^2$. The first value is               
the lowest mass expected based on the color-hyperfine interaction,          
while the latter value is that at which the largest number of               
events is observed in Fig.~\ref{fig:spectra}(a).                            
We consider mass windows of width 40~MeV/$c^2$ which, based                 
on the experimental resolution, should contain more than                    
90\% of true signal events.\\                                               
                                                                            
\noindent                                                                   
Table~\ref{tbl:ul} lists the numbers used in Eq.~(\ref{eqn:ratio}) and the 
resultant
upper limits. These limits include a small correction factor                
$f^{}_{\it SYS}$ to account for systematic                                  
uncertainties\cite{sys}. Such uncertainties                                 
are due mostly to differences                                               
between the simulation and data for                                         
the distributions of discrimination variables,                              
and their influence on the ratio                                            
$\varepsilon^{}_{P\rightarrow\phi \pi p} /                                  
     \varepsilon^{}_{D^{}_s\rightarrow\phi \pi}$.                           
The total systematic error is 19$\%$ for 
$M(P^0_{{\bar c}s})$=2.75 GeV/$c^2$, and 16$\%$ for 
$M(P^0_{{\bar c}s})$=2.86 GeV/$c^2$. These errors give rise to
correction factors $f^{}_{\it SYS}$=1.06 and $f^{}_{\it SYS}$=1.07,         
respectively.                                                               
Assuming the branching fractions of the $D^{\pm}_s\rightarrow\phi\pi^{\pm}$ 
and $P^0_{{\bar c}s}\rightarrow\phi\pi p$ decays are similar, the   
resulting upper-limits approach the range of the theoretically estimated    
ratio between the pentaquark                                                
and $D_s^{\pm}$ production cross sections~\cite{dpf94,danmur}.              
                                                                            
\begin{table}[ht]                                                          
\setlength{\baselineskip}{5.ex}                                            
  \centering                                                                
   \begin{tabular}{|c|c|c|}                                                 
M($P^0_{{\bar c}s})$ &  2.75  GeV/$c^2$  
                     &\multicolumn{1}{c|}{2.86  GeV/$c^2$}\\               
\hline                                                                      
$UL(N^{}_{P\rightarrow\phi\pi p})$ & 3.4 & \multicolumn{1}{c|}{11.0} \\    
$\varepsilon^{}_{P\rightarrow\phi\pi p}                                    
  / \varepsilon^{}_{D^{}_s\rightarrow\phi \pi}$&                           
           0.38$\pm$0.07 & \multicolumn{1}{c|}{0.62$\pm$0.09} \\           
\hline                                                                     
$N^{}_{D^{}_s\rightarrow\phi\pi}$   & \multicolumn{2}{c|}{293$\pm$18} \\   
\hline                                                                     
$90\%$ C.L.        &          & \multicolumn{1}{c|}{}        \\              
upper-limit        & 0.031    & \multicolumn{1}{c|}{0.063}  \\             
\end{tabular}                                                              
\caption{                                                                  
Values of $UL(N_{P\rightarrow\phi\pi p})$, ratio of efficiencies for       
$P\rightarrow \phi\pi p$ and $D_s\rightarrow\phi\pi$, and                  
$N_{D_s\rightarrow\phi\pi}$                                                
used to calculate the upper limit on the ratio of cross section times      
branching fraction for the decays $P^0\rightarrow\phi\pi p$ and            
$D_s^{\pm}\rightarrow \phi\pi^{\pm}$, as defined in Eq.~(1).               
Two values of pentaquark mass are used.                                    
The pentaquark lifetime used to calculate efficiencies is 0.4~ps.          
}                                                                          
\label{tbl:ul}                                                             
\end{table}                                                               
                                                                           
\noindent                                                                  
The value of the upper-limit depends upon the pentaquark lifetime due to   
dependence of the acceptance on lifetime. The upper-limit                  
is a rapidly decreasing function of lifetime,                              
from an upper-limit close                                                  
to 1 for 0.1 ps, to the value listed in the table for 0.4 ps, and remaining
about the same for larger lifetime values.\\

\noindent                                                                  
In summary,                                                                
we present results of the first search for the pentaquark particle.        
We see no convincing evidence for pentaquarks decaying to $\phi\pi p$      
in our data.                                                               
Upper-limits are presented for the ratio of $\sigma\cdot B$                
for $P^0_{{\bar c}s}\rightarrow\phi\pi p$ and                              
$D^{\pm}_s\rightarrow\phi\pi^{\pm}$, for two different values of 
pentaquark mass. The upper-limits are approaching                          
the theoretically estimated ratio of production cross-sections             
if we assume the same branching                                            
fraction for the two decays and a pentaquark lifetime of 0.4 ps or 
greater.\\

\noindent
We are grateful to Dr. Harry Lipkin for many fruitful discussions.         
We gratefully acknowledge the staffs of Fermilab and of all the            
participating institutions. This research was supported by the Brazilian   
Conselho Nacional de Desenvolvimento Cient\'\i fico e Technol\'ogio,       
the Mexican Consejo Nacional de Ciencia y Tecnologica,                     
the Israeli Academy of Sciences and Humanities, the U.S. Department        
of Energy, the U.S.-Israel Binational Science Foundation and the U.S. 
National Science Foundation. Fermilab is operated by the Universities 
Research Association, Inc., under contract with the United States 
Department of Energy.

\end{document}